\documentclass[ aps, prd, preprintnumbers,  floatfix, longbibliography, 10pt]{revtex4-2}

\usepackage{graphicx,amsfonts,amsmath,amssymb,amstext}
\usepackage{float,wrapfig}
\usepackage{subfigure, psfrag}
\usepackage{dsfont}
\usepackage{xcolor}
\usepackage[colorlinks=true,urlcolor=blue,citecolor=blue,linkcolor=blue]{hyperref}
\usepackage{bm}
\usepackage{mathtools}
\usepackage{enumitem}
\usepackage[normalem]{ulem}
\usepackage{cancel}

\graphicspath{{Figures/}} 

\definecolor{purple}{rgb}{0.8,0,0.6}
\definecolor{darkgreen}{rgb}{0.00,0.6,0.00}

\extrafloats{100}

\begin{document}

\title{Energy levels of screened charged impurity in gapped graphene under magnetic field}

\author{
Andrii Iurov$^{1}$\footnote{E-mail contact: aiurov@mec.cuny.edu, theorist.physics@gmail.com} and
V.P. Gusynin$^{2}$\footnote{E-mail contact: vgusynin@bitp.kyiv.ua}
}

\affiliation{
$^{1}$Department of Physics and Computer Science, Medgar Evers College of City University of New York, Brooklyn, NY 11225, USA \\
$^{2}$Bogolyubov Institute for Theoretical Physics, 14-b Metrolohichna Street, Kyiv 03143, Ukraine 
}

\date{\today}

\begin{abstract}
In this paper, we present a comprehensive theoretical study of the electronic states in gapped graphene in the presence of a charge impurity  and a magnetic field.
We focus on the impurity-induced modification of the Landau levels under the influence of an attractive or repulsive Coulomb potential.  Since analytical solutions 
of the Dirac equation are not available in this case, the study is carried out using the Wentzel–Kramers–Brillouin (WKB) approach. Special attention is paid to the  
screening of the Coulomb  potential in the presence of metallic gates situated at some distances from graphene sheet, as is commonly the case in experiments. Using 
both analytical and numerical methods, we determine  the energy spectrum as a function of the magnetic field strength and impurity charge for screened and unscreened 
Coulomb potentials.  Our results reveal a significant dependence of the energy levels for sample-to-gate distances of less than 50 nm.
\end{abstract}

\maketitle

\section{Introduction}
\label{sec1}

\medskip

Graphene-like two-dimensional materials  have attracted extensive interest because of their unusual electronic, optical, and collective properties  (for reviews see
\cite{Geim2007,Novoselov-SSC2007,IJMPA2007,CastroNeto-RMP2009,Peres2010,DasSarma-RMP2011}).
In monolayer graphene,  however, the absence of an intrinsic band gap limits conventional electronic applications and makes electrostatic confinement difficult because of Klein tunneling \cite{Katsnelson2006,Beenakker2008,Allain2011}. Opening a gap transforms the massless Dirac spectrum into a massive one and substantially modifies the electronic structure, density
 of states, transport, and confinement properties. Gapped graphene can be realized through sublattice-symmetry breaking by substrates \cite{Giovannetti2007}, using ribbons \cite{n11},
chemical adsorption \cite{Lu2012}, or external fields and interactions \cite{Khveshchenko2001,Gorbar2002,Herbut2006}. The finite gap suppresses Klein tunneling  enabling confinement
in electrostatic and magnetic structures \cite{n14}. These changes have important consequences for collective properties, such as an extended plasmon mode \cite{n15},\cite{n16}.

The spectrum of a single unscreened charged impurity was in fact known from the study of the massive Dirac equation in two spatial dimensions in the Coulomb field, for which—as
in the three-dimensiona case—exact solutions are known \cite{Khalilov1998} (rederived later in \cite{Reity2002,Dong2002}). In the case of monolayer gapped graphene a similar
bound-state hydrogene-like spectrum was reproduced in the framework of the low-energy continuum description in \cite{Novikov2007,Gupta2008,Gamayun2009}. The theoretical
prediction \cite{Pereira2007,Shytov1,Shytov2} and experimental confirmation \cite{Wang2013,Mao2016} of the atomic-collapse effect in gapless graphene has attracted particular
attention further confirming that graphene can serve as a desktop laboratory for testing predictions of quantum electrodynamics (QED). Whereas in the real QED, this effect is associated
with the production of positrons for large supercritical charges of the Coulomb center, $Z>Z_c\sim170$ \cite{Zeldovich1971,Popov2001}, which are not observed in nature, —in
graphene it manifests as the appearance of sharp peaks (electronic resonances) in the local density of states at supercritical charges. Because charge carriers in graphene have a large 
``fine-structure constant'' $\alpha$, the supercritical charge is substantially lower than for atoms, with $Z_c$ on the order of unity. The phenomenon of atomic collapse is tightly related
to excitonic instability in gapless graphene \cite{Gamayun2009,Gamayun2010}, quasiparticle gap generation and the phase transition semimetal--insulator  when the long-range, Coulomb
interaction between 2D massless Dirac fermions exceeds a critical threshold ($\alpha>\alpha_c$)
\cite{Gorbar2002,Leal2004,Drut2009,Liu2009,Gamayun2010, Sabio2010,Fertig2011,Buividovich2012,Popovici2013}, this is similar to chiral  symmetry breaking in QED.
However,  spontaneous semimetal-insulator transition is not realized in real, pristine  suspended monolayer graphene, which remains in a conducting (semimetallic) phase. This is because
in clean suspended graphene the Fermi velocity increases when close to the charge neutrality point, thus making electron-electron Coulomb interaction weaker \cite{Juan2010,Elias2011}.

Graphene exhibits rich and unique magnetic-field phenomena and properties arising from its two-dimensional Dirac spectrum, with a perpendicular magnetic field quantizing the electronic states
 into relativistic Landau levels.  Overall, magnetic fields provide one of the most powerful tools for probing and controlling the electronic, transport, optical, and many-body properties of graphene \cite{IJMPA2007,CastroNeto-RMP2009,Peres2010,DasSarma-RMP2011}. It also significantly alters the graphene spectrum by transforming the valence and  conduction band into discrete
 Landau levels, allowing for bound states even under a repulsive interaction. A significant difference from the  purely Coulombic case is that,  in the presence of a magnetic field.  the Dirac equation
 with a Coulomb potential cannot be solved in terms of known special functions; analytic solutions are possible for large magnetic fields for a countably  infinite set of magnetic
 field strengths \cite{Khalilov2000,Chiang2002}. The standard methods for solving the Dirac equation in this case are either numerical calculations \cite{ Sun2014,Pottelberge2018}, or the
 use of approximate  analytical methods, primarily the  WKB method. For example, the WKB method in the case of gapless graphene was applied  in Refs.\cite{Carmier2008,Zhang2012, Reijnders2013,Nimyi2022}.
 In particular, gapless  graphene was studied in the presence of a magnetic field, perpendicular to the graphene plane, and a charged Coulomb center \cite{Zhang2012},  as well a radial constant
 electric field \cite{Nimyi2022}. In both cases, the Bohr–Sommerfeld integral is calculated exactly and expressed in terms of complete elliptic integral functions. It is important to note a key
 difference between the WKB method for systems of linear differential equations and scalar equations of the Schrödinger type. For the system of equations, the WKB ansatz
 $\Psi(x)=A(x)e^{i S(x)/\hbar}$ requires an amplitude $A(x)$ to be a vector, and a change of the vector orientation along a path leads to the appearance of a geometric phase (such as the Berry
 phase) \cite{Littlejohn}. The latter is a most distinctive feature of the WKB method for systems. For applications to other Dirac-like condensed matter systems see, for example, \cite{Iurov2021,IurovWKBBook2024}.

 It is also worth noting the exact numerical studies of a Coulomb center  in the honeycomb lattice within the framework of tight-binding models,
 or by diagonalization of the large Hamiltonian matrices, which consider phenomena like bound states, atomic collapse without and with a magnetic field, and compare the results to continuum
 Dirac equation models  \cite{Pereira2008,Zhu2009, Kim2014, Moldovan2018}.  In particular, anticrossings of levels (level repulsions) for supercritical charge described in \cite{Sobol2016}
 was confirmed by numerical studies using a tight-binding approach \cite{Moldovan2018}.

In this paper, we study bound states in gapped graphene in the presence of a perpendicular magnetic field and a radially symmetric potential $V(r)$. We model $V(r)$ as a Coulomb
potential screened by polarization effects and  the presence of metallic gates (see, e.g., \cite{Shklovskii2024}). This  screening is the main factor that distinguishes our work from
previous studies. Taking into account screening caused by nearby metallic gates may be important when analyzing experimental results. It is shown also that in the WKB approach
to the Dirac equation the contribution of the geometric phase  is described by an integral proportional to a gap and the derivative of the potential $V'(r)$ and plays an important role for
calculating energy levels. Since the potential itself is determined by a certain integral, we derive a fitting analytical formula to calculate the Bohr–Sommerfeld  and
the geometric phase integrals. Our numerical results for energy levels of a charged impurity under external magnetic field  are presented in a series of figures.

\medskip

The  paper is organized as follows: Section \ref{sec2} presents a detailed Wentzel–Kramers–Brillouin investigation of the Dirac equation for gapped graphene in a magnetic field
with radially symmetric interaction potential. We apply the WKB approximation to obtain wave functions and a generalized Bohr-Sommerfeld quantization condition, which takes
into account geometric phase, for graphene in a magnetic field in the presence of a charged impurity. We also considered the limiting cases of our problem: vanishing Coulomb
interaction -- regular Landau levels in magnetic field, and zero magnetic field, which corresponds to an electron in an electrostatic Coulomb trapping potential.
The  derivation for the latter case is provided in Appendix \ref{apaa} which demonstrates the role of geometric phase for obtaining exact energy levels from the WKB
solutions in this particular case.
Next, we introduce a screened Coulomb potential in Section \ref{sec3}, which takes into account the electron-electron polarization function and screening due to metallic gates situated
at some distances above and below graphene plane.    In particular, we present an analytical fitting model for this potential which  helps compute the Bohr-Sommerfeld integral.
Our numerical results and  the corresponding discussion are presented in Sec.~\ref{sec4}.
We analyze how Landau levels depend on the impurity charge of both signs at fixed magnetic field, and
for fixed impurity  charge and the varying magnetic field, and provide the detailed results for Landau levels of gapped graphene in the presence of the screened potential.
Several useful relations for the screened potential and its various analytical approximations are presented in Appendix \ref{apb}. Finally,  we provide our concluding remarks in Section \ref{sec5}.

\section{WKB expansion for gapped graphene with radially symmetric potentials}
\label{sec2}
The electron quasiparticle states in gapped graphene with  a magnetic field and Coulomb-like impurity are described  at the $K_\xi$ point of the Brillouin zone  by the low-energy
Dirac Hamiltonian
\begin{align}
H_\xi= v_F\boldsymbol{\sigma}\boldsymbol{\pi}+\xi\Delta\sigma_3+V(r)=\left(\begin{array}{cc}
\xi\Delta +V(r)&v_F{\pi_-}\\ v_F{\pi_+}&-\xi\Delta+V(r)\end{array}\right),
\label{hamiltonian-gapped_graphene}
\end{align}
where $v_F\approx10^6$ m/s is the Fermi velocity, $\sigma_i$ are the Pauli matrices, $\boldsymbol{\pi}=-i\hbar\boldsymbol{\nabla}+(e/c)\mathbf{A}$ is the canonical momentum
with the vector potential in the symmetric gauge $\mathbf{A}=(B/2)(-y,x)$ corresponding to the magnetic field strength $B>0$ in $z$-direction, $\pi_\pm=\pi_x\pm i\pi_y$ and $-e<0$
is the electron charge. As for  charged impurity potential $V(r)$, we will consider as the pure Coulomb potential $\sim 1/r$ as well as potentials modified due to the presence of polarization
effects and screening by the presence of metallic gates. The  Coulomb potential is $V_C(r)=\hbar v_Fg/r$ where $g=Z\alpha$ in which $Z$ is  the impurity charge, $\alpha=e^2/(\kappa\hbar v_F)$
is graphene’s fine structure constant, and $\kappa$ is the effective dielectric constant (for example, for SiO$_2$ substrate $\kappa\approx2.4$). We included also a gap parameter $\Delta$ in
Eq.(\ref{hamiltonian-gapped_graphene}) where $\xi=\pm$ is the valley index denoting the Dirac $K_\pm$ points.

A measurable bandgap $2\Delta$ can be induced in graphene by an underlying dielectric substrate, with its magnitude depending strongly on the substrate and graphene–substrate
alignment. ARPES measurements  of epitaxial graphene on SiC reported a gap of approximately $0.26$ eV \cite{ZhouSubstrate2007}, while first-principles calculations predicted
a gap of about $53$ meV for graphene on h-BN due to sublattice symmetry breaking \cite{Giovannetti2007}. We use $\Delta=53$ meV in our calculations, consistent  with
 experimentally reported gaps of approximately $20$--$100$ meV in graphene/h-BN heterostructures \cite{HuntMassive2013, WangGaps2016, HanAccurate2021}.
 A bandgap up to
 $160$ meV has been theoretically predicted  single-layer graphene sandwiched between two h-BN layers, \cite{Quhe2012Tunable} therefore, there exists a good tunability of the gap parameter.
\medskip

The eigenfunctions can be written in polar coordinates as
\begin{align}
\Psi_j(r,\phi)=\frac{1}{\sqrt{r}}\left(\begin{array}{c}F_j(r)e^{i(j-1/2)\phi}\\ i G_j(r)e^{i(j+1/2)\phi}\end{array}\right),
\end{align}
where $j$ is an eigenvalue of the total angular momentum $J_z=L_z+S_z=-i\hbar\partial_\phi +\hbar\sigma_z/2$ and takes half-integer values
$j=\pm 1/2,\pm 3/2,\dots$ (in what follows for simplicity we omit subscript $j$ in functions $F_j$ and $G_j$).

 Hence for radially symmetric potentials the Dirac equation can be represented in the form
 \begin{align}
\hbar \frac{d}{d r}\Phi(r)=D\Phi(r),
\label{Dirac_radial}
\end{align}
where $\Phi(r)=(F(r),G(r))^T$ is a spinor function and the matrix $D$ is
\begin{align}
D=\left(\begin{array}{cc}\frac{J}{r}+\frac{e B r}{2c}& -\left(\frac{E+\Delta-V(r)}{v_F}\right)\\
\frac{E-\Delta-V(r)}{v_F}&-\left(\frac{J}{r}+\frac{e B r}{2c} \right) \end{array}\right).
\end{align}
 In the WKB expansion the energy and total angular momentum  $J=\hbar j$ are treated as independent of $\hbar$ since they are conserved physical quantities, hence
 the matrix $D$ does not depend on $\hbar$. Within the WKB expansion, we expand the solution of Eq. (\ref{Dirac_radial}) in the form
\begin{align}
\Phi(r)=e^{i S(r)/\hbar}\psi(r),\quad \psi(r)=\sum\limits_{n=0}^\infty (-i\hbar)^n \psi^{(n)}(r).
\end{align}
The equation for $\psi(r)$ takes the form
\begin{align}
(M_0+\hbar M_1)\psi=0,
\label{eq:Psi}
\end{align}
where the matrices $M_0,M_1$ are
\begin{align}
M_0=\left(\begin{array}{cc}a(r)-iS'(r)&-b(r)\\ b_1(r)&-a(r)-iS'(r) \end{array}\right),\qquad M_1=-\left(\begin{array}{cc}1&0\\0&1
\end{array}\right)\frac{d}{dr}.
\end{align}
Here we introduced the notations
\begin{align}
a(r)=\frac{J}{r}+\frac{e B r}{2c},\quad b(r)=\frac{E+\Delta-V(r)}{v_F},\quad b_1(r)=\frac{E-\Delta-V(r)}{v_F}.
\label{def:a,b}
\end{align}
In the lowest order approximation in $\hbar$ nontrivial solution for $\psi^{(0)}$ exists if $det M_0=0$ which gives
\begin{align}
S'_{\pm}(r)=\pm p(r),\qquad p(r)=\sqrt{b(r)b_1(r)-a^2(r)}=\sqrt{\frac{(E-V)^2-\Delta^2}{v_F^2}-
\left(\frac{J}{r}+\frac{e  B r}{2c}\right)^2}.
\end{align}
Right and left eigenvectors with zero eigenvalue of the matrix $M_0$, in which  $S'_\pm (r)=\pm p(r)$, are defined as $M_0\psi_{\pm}^{(0)}(r)=0$ and $\tilde{\psi}_{\pm}^{(0)}(r)M_0=0$, respectively.
They can be written in two equivalent but different forms
\begin{align}
&\psi_{\pm}^{(0)}(r)=C^{\pm}_R(r)\left(\begin{array}{c}b(r)\\a(r)\mp ip(r)\end{array}\right)=D^{\pm}_R(r)\left(\begin{array}{c}
a(r)\pm i p(r)\\b_1(r)\end{array}\right),\\
&\tilde{\psi}_{\pm}^{(0)}(r)=C^{\pm}_L(r)\left(-b_1(r), a(r)\mp ip(r)\right)\,=\, D^{\pm}_L(r)\left( a(r)\pm ip(r),-b(r)\right).
\label{right_spinors}
\end{align}
Here $C^{\pm}_{L,R}(r)$, $D^{\pm}_{L,R}(r)$ are normalization functions. Substituting now  the expression for $\psi_{\pm}^{(0)}(r)$ into Eq.(\ref{eq:Psi})  we  get the equation
for the functions $C^{\pm}_R(r)$:
\begin{align}
\frac{\partial_r C^{\pm}_R}{C^{\pm}_R}=-\frac{1}{2}\frac{d}{dr}\log[-b b_1+(a\mp ip)^2]+\frac{1}{2}\frac{b b'_1-b'b_1}{-b b_1+(a\mp ip)^2},
\end{align}
whose solution is given by the expression
\begin{align}
C^{\pm}_R(r)= \frac{\tilde{C}^{\pm}}{[p(r)(p(r)\pm ia(r))]^{1/2}}\exp\left(-\int^rdr\frac{\Delta V'(r)}{2v_F^2p(r)(p(r)\pm ia(r))}\right),
\end{align}
with $\tilde{C}^{\pm}$ being constants. Thus we have the WKB solutions for right vectors in the form
\begin{align}
\psi^{(0)}_{\pm}(r)=\frac{\tilde{C}^{\pm}}{[p(r)(p(r)\pm ia(r))]^{1/2}}\exp\left(\pm (i/\hbar)\int^rp(r)dr+\Gamma(r)\right)
\left(\begin{array}{c}b(r)\\a(r)\mp ip(r)\end{array}\right),
\label{gapped_graphene-WKB_solutions}
\end{align}
where
\begin{align}
\Gamma(r)=-\int^r\frac{\Delta V'(r)dr}{2v_F^2p(r)(p(r)\pm ia(r))}.
\end{align}
The last expression can be written in the form
\begin{align}
\Gamma(r)=-\frac{1}{4}\log\frac{E-V(r)+\Delta}{E-V(r)-\Delta}\pm i\gamma(r)=-\frac{1}{4}\log\frac{b(r)}{b_1(r)}\pm i\gamma(r),
\end{align}
where $\gamma(r)$ is the phase function,
\begin{align}
\gamma(r)=\frac{\Delta}{2}\int^r\frac{dr V'(r)a(r)}{p(r)[(E-V(r))^2-\Delta^2]}.
\end{align}
To extract phases from spinors we write
\begin{align}
p(r)\pm ia(r)=\sqrt{p^2(r)+a^2(r)}e^{\pm 2i\varphi(r)},\qquad \varphi(r)=\frac{1}{2}\arctan\frac{a(r)}{p(r)}
\end{align}
and
\begin{align}
p^2(r)+a^2(r)=b(r)b_1(r)=\frac{(E-V(r))^2-\Delta^2}{v_F^2}.
\label{nearexponential_factor}
\end{align}
The solutions (\ref{gapped_graphene-WKB_solutions}) can be finally written in the form
\begin{align}
\psi^{(0)}_{\pm}(r)=\frac{\tilde{C}^{\pm}}{[p(r)|b(r)|]^{1/2}}\exp\left(\pm (i/\hbar)\int^rp(r)dr \pm i\gamma(r)\right)
\left(\begin{array}{c}b(r)e^{\mp i\varphi(r)}\\ \mp i\sqrt{b(r)b_1(r)}\, e^{\pm i\varphi(r)}\end{array}\right).
\label{Psi-WKB-nonzeroDelta}
\end{align}
We can obtain slightly different form for the spinors $\psi^{(0)}_{\pm}(r)$ using the second form for the right spinor in Eq.(\ref{right_spinors}),
\begin{align}
\psi^{(0)}_{\pm}(r)=\frac{\tilde{D}^{\pm}}{[-p(r)(p(r)\mp ia(r))]^{1/2}}\exp\left(\pm (i/\hbar)\int^rp(r)dr+\int^r\frac{\Delta V'(r)dr}
{2v_F^2p(r)(p(r)\mp ia(r))}\right)\left(\begin{array}{c}a(r)\pm ip(r)\\ b_1(r)\end{array}\right).
\label{gapped_graphene-WKB_solutions-2}
\end{align}
Different forms are used in order to avoid a singularity of the wave function at the point where $b(r)=0$ or $b_1(r)=0$ (see Eq.\ref{nearexponential_factor}), when
we are in classically admissible region. Moreover, in classically forbidden regions, where after analytic continuation $p(r)\rightarrow \pm iq(r)$,
the use of forms depends on the sign of $a(r)$ to avoid singularity.

Using the single-valuedness condition for components of the wave function, for example, $\psi^{(0)}_+(r)$, we obtain the Bohr-Sommerfeld like quantization condition
\begin{align}
\frac{1}{\hbar}\int\limits_{r_-}^{r_+}dr p(r)+\frac{\Delta}{2}\int\limits_{r_-}^{r_+}\frac{dr V'(r)a(r)}{p(r)[(E-V(r))^2-\Delta^2]}
\mp\frac{1}{2}\int\limits_{r_-}^{r_+}dr\left(\arctan\frac{a(r)}{p(r)}\right)'=\pi\left(n_{BS}+\frac{1}{2}\right),\quad n_{BS}=0,1,\dots,
\label{WKB-rule-Delta}
\end{align}
where signs $\mp$ correspond to the signs of phases in upper and lower components of the spinor $\psi^{(0)}_+(r)$ in
Eq.(\ref{Psi-WKB-nonzeroDelta}) and the turning points $r_{\pm}$ are determined by the equation $p^2(r)=0$. Taking into account the
definition of $a(r)$ in Eq.(\ref{def:a,b}) we find that
\begin{align}
\int\limits_{r_-}^{r_+}dr\left(\arctan\frac{a(r)}{p(r)}\right)'=\pi\theta(-j).
\end{align}
We recall that the function $a(r)$ in Eq.(\ref{def:a,b}) is positive for positive $j$, hence the last integral vanishes for $j>0$. For negative $j$, $a(r)$
is monotonically increasing function, which changes sign at the point $r_0 = l\sqrt{2j}$ in the classically allowed region. Thus the quantization
condition (\ref{WKB-rule-Delta}) takes the final form
\begin{align}
\frac{1}{\hbar}\int\limits_{r_-}^{r_+}dr p(r)+\frac{\Delta}{2}\int\limits_{r_-}^{r_+}\frac{dr V'(r)a(r)}{p(r)[(E-V(r))^2-\Delta^2]}
=\pi\left(n_{BS}+\frac{1\pm\theta(-j)}{2 }\right),\quad n_{BS}=0,1,\dots.
\label{WKB-rule-Delta_final}
\end{align}
In the last expression we must take the series with the minus sign before $\theta(-j)$, since it already includes the series with the plus sign before $\theta(-j)$.
\vskip3mm
Thus we have final formula for WKB energy levels in a potential $V(r)$ in the presence of a magnetic field
\begin{align}
\frac{1}{\hbar}\int\limits_{r_-}^{r_+}dr p(r)+\frac{\Delta}{2}\int\limits_{r_-}^{r_+}\frac{dr V'(r)a(r)}{p(r)[(E-V(r))^2-\Delta^2]}
=\pi\left(n_{BS}+\frac{1-\theta(-j)}{2 }\right),\quad n_{BS}=0,1,\dots,
\label{WKB-Delta+magfield_final}
\end{align}
where the local momentum is given by
\begin{align}
p(r)=\sqrt{\frac{(E-V(r))^2-\Delta^2}{v_F^2}-\left(\frac{\hbar j}{r}+\frac{\hbar r}{2l^2}\right)^2},
\end{align}
and
\begin{equation}
l = \sqrt{\frac{\hbar c}{e B}}  \,\, \rightarrow \,\, l(nm) = \frac{25.66\,\text{nm}}{\sqrt{B\text{(T)}}}
\end{equation}
is the magnetic length in nm as a function of $B$ in T. For all our calculations with a fixed magnetic field (and fixed magnetic length) and varying impurity charge $g$, we use
$l = 12.38$nm, which corresponds to $B=4.30$T.

\medskip
In the case of the Coulomb potential $V_C(r)=-Ze^2/(\kappa r)=-\hbar v_F g/r$ it is convenient to introduce dimensionless variables
\begin{align}
\rho=\frac{r}{l},\quad \epsilon=\frac{l E}{\hbar v_F},\quad \delta=\frac{l \Delta}{\hbar v_F},
\end{align}
and rewrite Eq.(\ref{WKB-Delta+magfield_final}) in the form
\begin{align}
\int\limits_{\rho_-}^{\rho_+} d\rho\tilde{p}(\rho)+\frac{g\delta}{2}\int\limits_{\rho_-}^{\rho_+ }\frac{d\rho\tilde{a}(\rho)}
{\rho^2\tilde{p}(\rho)[(\epsilon+g/\rho)^2-\delta^2]}=\pi\left(n_{BS}+\frac{1-\theta(-j)}{2 }\right),\quad n_{BS}=0,1,\dots,
\label{eq:quntization}
\end{align}
where
\begin{align}
\tilde{p}(\rho)=\sqrt{(\epsilon+g/\rho)^2-\delta^2-\tilde{a}^2(\rho)},\qquad \tilde{a}(\rho)=\frac{j}{\rho}+\frac{\rho}{2},
\end{align}
and $\rho_{\pm}$ are the roots of the equation $\tilde{p}^2(\rho)=0$. In more explicit form Eq.(\ref{eq:quntization}) reads
\begin{align}
&\int\limits_{\rho_-}^{\rho_+} \frac{d\rho}{\rho}\sqrt{(\epsilon\rho+g)^2-\delta^2\rho^2-(j+\rho^2/2)^2}+\frac{g\delta}{2}\int\limits_{\rho_-}^{\rho_+ }
\frac{d\rho(j+\rho^2/2)}{\sqrt{(\epsilon\rho+g)^2-\delta^2\rho^2-(j+\rho^2/2)^2}[(\epsilon\rho+g)^2-\delta^2\rho^2]}\nonumber\\
&=\pi\left(n_{BS}+\frac{1-\theta(-j)}{2 }\right),\quad n_{BS}=0,1,\dots\,.
\label{BS:Coulomb+magfield}
\end{align}
This dimensionless form of the energy levels equation is subsequently used for numerical calculations. We now consider two limiting cases, $g=0$ and $B=0$.

\medskip
\subsection{WKB energy levels in a magnetic field without Coulomb impurity}
In the absence of Coulomb potential ($g=0$) Eq.(\ref{BS:Coulomb+magfield}) takes the form
\begin{align}
\int\limits_{\rho_-}^{\rho_+} \frac{d\rho}{\rho}\sqrt{(\epsilon^2-\delta^2)\rho^2-(j+\rho^2/2)^2}=\pi\left(n_{BS}+\frac{1-\theta(-j)}{2 }\right),\quad n_{BS}=0,1,\dots\,.
\label{BS:magfield}
\end{align}
Obviously, solutions exist only for $|\epsilon|>\delta$. The integral is easily calculated and we obtain
\begin{align}
\epsilon^2=2n_{BS}+\delta^2+j+|j|+1-\theta(-j),\qquad n_{BS}=0,1,\dots\,.
\end{align}
For positive $j\ge1/2$ we have
\begin{align}
\epsilon=\pm\sqrt{2(n_{BS}+j+1/2)+\delta^2}.
\end{align}
Introducing an integer number $n=n_{BS}+j+1/2$ we  reproduce the exact Landau spectrum of Dirac particle in a perpendicular  constant magnetic field $\epsilon=\pm\sqrt{2n+\delta^2}$
where $n=1,2,\dots$ and $1/2\le j\le n-1/2$. For negative $j\leq-1/2$ we obtain $\epsilon=\pm\sqrt{2n_{BS}+\delta^2}$ and we identify $n_{BS}=n$, $n=0,1,\dots$, as the Landau level index.
The zeroth Landau level corresponds $n_{BS}=n=0$ and $\epsilon=\pm\delta$. However, in this case the two positive real roots of $p^2(r) = 0$ equal to each other, and the integral in
Eq.(\ref{BS:magfield}) is zero. Thus the WKB method reproduces well-known exact results except zeroth Landau level (see, for example, \cite{Zhang2012}). In the exact solution the
zeroth Landau level at the $K_-$-point is $\epsilon=-\delta$, the only level which makes the energy spectrum asymmetric with respect to the point $\epsilon=0$.
At the $K_+$-point the zeroth Landau level is at $\epsilon=\delta$ and combined spectrum at two Dirac points is symmetric \cite{IJMPA2007}.

\subsection{Zero magnetic field and non-vanishing Coulomb potential}.
\label{apa}

Let us show that in  the case of attractive Coulomb potential $V_C(r)=-\hbar v_F g/r$ without magnetic field  the WKB quantization condition  gives energy levels coinciding
with known exact solutions of ($2+1$)-dimensional Dirac equation. The integral with arctan in Eq.(\ref{WKB-rule-Delta}) vanishes in this case and for the second integral in
the left hand side
we have
\begin{align}
I=\frac{\Delta}{2}\int\limits_{r_-}^{r_+}\frac{dr V'(r)a(r)}{p(r)[(E-V(r))^2-\Delta^2]}=\frac{\Delta g(\hbar v_F)^2 j}{2}\int\limits_{r_-}^{r_+}
\frac{dr}{\sqrt{(E r+g\hbar v_F)^2-\Delta^2r^2-(\hbar v_F j)^2}[(E r+g\hbar v_F)^2-\Delta^2r^2]}.
\end{align}
This integral can be evaluated exactly and in Appendix \ref{apaa} we find  the energy levels in the Coulomb potential
\begin{align}
\epsilon_{n,j}=\left[1+\frac{g^2}{(n+\sqrt{j^2-g^2})^2}\right]^{-1/2},\quad\left\{\begin{array}{c}n=0,1,\dots,\,j>0,\\
n=1,2,\dots,\,j<0.\end{array}\right.
\label{Coulomb-energies}
\end{align}
The last formula is in exact agreement with Eq.(13) in \cite{Khalilov1998}
 (this result was rederived in  \cite{Reity2002,Dong2002} and for gapped graphene in \cite{Novikov2007, Gupta2008, Gamayun2009}).
The results are valid for $g<1/2$ and all $j=\pm 1/2, \pm 3/2,\dots$\,. The bound states are doubly degenerate, $\epsilon_{n,j}=\epsilon_{n,-j}$, except the ground state with $n=0, j=1/2$.

\section{Screened potential of charged impurity}
\label{sec3}

The Coulomb potential of impurity screened by internal electron polarization and by tunable gates was obtained in Ref.\cite{Shklovskii2024}. The interaction energy of electron
with impurity of a charge $q$ in the presence of internal static polarization $\Pi(k)$ and upper and lower gates, both  at distances $d$ from graphene plane, is given by the expression
\begin{align}
V(r)=-\frac{e q}{\kappa}\int\limits_0^\infty\frac{dk k J_0(k r)}{k\coth(kd)+\frac{2\pi e^2}{\kappa}\Pi(k)},
\label{pot-two_gates_equal}
\end{align}
where  $J_0(k)$ is the Bessel function and $\kappa$ is the dielectric constant of a medium between gates and graphene, for example, for hexagonal
boron nitride (hBN) $\kappa\approx4$. The distance $d$ can be varied from 10 to 200 nm.

There is a more general formula when the upper gate is at distance $d_1$ and the lower gate is at other distance $d_2$ from graphene  with dielectric
constants $\kappa_1$ and $\kappa_2$ of media, respectively, it has the form
\begin{align}
V(r)=-2e q\int\limits_0^\infty\frac{dk k J_0(k r)}{k\kappa_1\coth(kd_1) +k\kappa_2\coth(kd_2) +4\pi e^2 \Pi(k)},
\label{pot-two_gates_diff}
\end{align}
which obviously reduces to Eq.(\ref{pot-two_gates_equal}) when $d_1=d_2=d$ and $\kappa_1=\kappa_2=\kappa$. The last formula allows to
study the case of one gate. For example, taking the limit $d_2\to\infty$, we obtain ($\kappa_2=1$)
\begin{align}
V(r)=-2e q\int\limits_0^\infty\frac{dk k J_0(k r)}{k\kappa_1\coth(kd_1) +k +4\pi e^2 \Pi(k)}.
\label{pot-one_gate}
\end{align}

The above interaction energies do not include screening due to core electrons related to $\sigma$ bonds. It is known from studies in single \cite{Ulybyshev2013} and bilayer
graphene \cite{Santos2013} that core electrons affect the polarization function. This contribution can be accounted by replacement $2\pi e^2\Pi(k)$ in  Eq.(\ref{pot-two_gates_equal})
with  $2\pi e^2\Pi(k)+\sigma_gk$. In graphene $\sigma_g=0.9$ that gives dielectric constant $\epsilon \approx 1.4$ \cite{Santos2013}, which defines screening of the bare Coulomb
potential in suspended graphene ($\kappa = 1$) and in the absence of gates ($d \to\infty$) and vanishing polarization $\Pi(k)$ due to itinerant electrons.

\medskip
The polarization function in graphene was calculated in Ref.\cite{Gorbar2002}, for zero doping it is given by Eq.(14) there which we adapt
to definitions used here ($\epsilon_0=2\pi$, and $N_f=2$ for graphene) and restoring $\hbar$:
\begin{align}
2\pi e^2\Pi({\bf k})&=\frac{2e^2}{(\hbar v_F)^2}\left[\Delta+ \frac{(\hbar v_F)^2{\bf k}^2-4\Delta^2}{2\hbar v_F|{\bf k}|}
\arctan\frac{\hbar v_F|{\bf k}|}{2\Delta}\right]\nonumber\\
&=2\alpha\left[\frac{\Delta}{\hbar v_F}+\left(\frac{k}{2}-\frac{2\Delta^2}{(\hbar v_F)^2k}\right)\arctan\frac{\hbar v_F k}{2\Delta}\right],\qquad
k=|{\bf k}|,
\end{align}
and $\alpha=e^2/\hbar v_F\approx2.19$ is already introduced the  fine-structure constant of graphene. The characteristic length related to a gap $\Delta$ is $\lambda=\hbar v_F/\Delta$.
Since $\hbar\approx 6.6\cdot 10^{-16}eV\cdot s$ and taking $v_F\approx10^6$m/s, we get  $\lambda={0.66nm}/{\Delta(eV)} $.

Asymptotic behavior of the polarization function is given by the formulas
\begin{align}
2\pi e^2\Pi({\bf k})=\left\{\begin{array}{c}2\alpha\frac{\hbar v_F k^2}{3\Delta},\quad k\ll \Delta/\hbar v_F,\\
\pi\alpha  k/2,\qquad k\gg \Delta/\hbar v_F\end{array}\right.\,.
\end{align}
As can be seen from these formulas, the quantity $\Delta$  serves as a threshold value below which small wave vectors $\mathbf{k}$  (long distances) are suppressed.
Evidently,  as $\Delta$ increases, the role of polarization decreases significantly.

In our numerical computations of the WKB integrals  we use the following analytical fitting expression  for the screened potential (\ref{pot-two_gates_equal}),

\begin{align}
\label{fitP01}
V_{\text{fit}}(r,d) = - g \, c \, \left\{ \frac{a}{r} + \frac{b}{\sqrt{r d}} \right\} \, \left[ 1 - e^{- f  d} \right] \, ,
\end{align}
where the lengths $r$ and $d$ are now given in terms of the magnetic length $l$ and the potential itself in terms of the energy scale $\hbar v_F/l$. This fitting takes into
account that for large distances $d$ between a sample and gates (ideally, $d \rightarrow \infty$) this interaction becomes equivalent to a regular Coulomb interaction
$\backsim g/r$. The distinction between the two expressions is exponentially reduced for an increasing $d$. For a smaller $d$, the $r$-dependence of our effective
potential is represented by a Coulomb term $g (a \, c)/r$ and a weaker dependence $b/\sqrt{r d}$, which also disappears for $d \rightarrow \infty$.

\par
To find an analytical fit for the screened potential, we applied an optimization procedure with 500 randomly distributed plot points shown in Fig.~\ref{FIG:4} $(b)$. We have obtained the
following values for the coefficients of Eq.~\eqref{fitP01}: $a = 0.06612$, $b = - 0.016116$, $c = 12.3735$ ($a c = 0.81813$; in the absence of the polarization it must be $1.000$ and
we have obtained $a  c = 1.00264$), and $f = 0.89368$.

We  compared  analytical fitting model \eqref{fitP01} with the numerically calculated potential \eqref{pot-two_gates_equal}.  As far as the $d$-dependence
is concerned, the analytical model agrees very well with the numerical results over most of its relevant range, with noticeable deviations only for very small values, $d  \ll 1$. However,
such small values of $d$ are not  practically relevant for gate distances, and when $d$ increases, the agreement becomes stronger, approaching complete coincidence in the limit
$d \rightarrow \infty$.
Regarding the $r$-dependence, the analytical model also shows good agreement with the numerical results over nearly the entire range, with deviations occurring only at very small and very
large $r$. Importantly, our model requires the $r$-dependence only between the turning points, which mostly span the range $1.0<r <7.0$ (in dimensionless units). Within this physically
relevant interval, the analytical fitting model provides very good agreement with the numerical results.

   The fitting \eqref{fitP01} was used for calculations with a fixed magnetic field, but it can also be used to study magnetic-field-dependent behavior. In this case, the lengths are made
   dimensionless with respect to the Compton length $\lambda$, and the potential with $\Delta$. In our calculations, we fixed the value of $\Delta=53$ meV to ensure consistency with certain
   experimental data; therefore, to ensure that the fitting is applicable to both cases, a fixed magnetic field and  a varying magnetic field, with the same numerical values of the parameters
   $a$, $b$, $c$, and $f$, it is convenient to choose the normalization lengths $l_0 = \lambda_0$, hence our choice of $l_0=12.38$ nm.

\section{Results and discussions}
\label{sec4}

Let us now briefly describe the system of units we use to present our numerical results.  The fixed energy bandgap for graphene which is equal to $2 \Delta$, where
$\Delta = 53$ meV, is taken as our unit of energy.  The unit of electron wave vector is $k_0 = \Delta/(\hbar v_F) =8.08\cdot 10^{7}$m$^{-1}$, and its reciprocal
$\lambda_0 = \hbar v_F/\Delta = 12.38$ nm ($=l_0$) is our unit of length.

\begin{figure}
\centering
\includegraphics[width=0.9\textwidth]{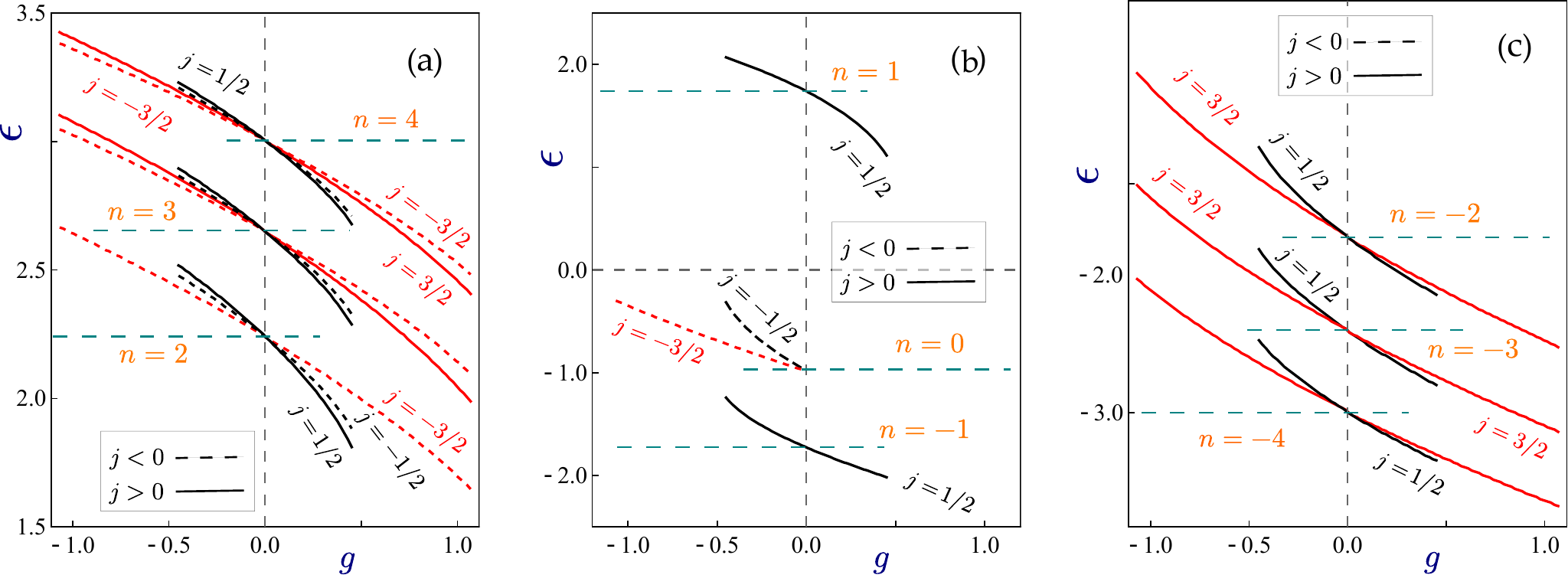}
\caption{(Color online)
The energy levels -- distorted Landau levels -- of graphene with a finite bandgap $\Delta=53$ meV in a uniform magnetic field in the presence of
a Coulomb impurity as a function of its charge $g$ with a zero out-of-plane distance $d_\perp=0$ to the impurity.  Panels (a), (b) and (c) show the energy levels with quantum
numbers $4\ge n\ge-4$  and angular momenta $j = \pm1/2 $  (black curves) and $j = \pm3/2$  (red curves), as labeled. The dotted horizontal lines correspond to the energy levels
$\epsilon(n)={\rm sign}(n)\sqrt{2|n|+1}$ in a magnetic field in the absence of a charged impurity.
}
\label{FIG:1}
\end{figure}
\medskip
\begin{figure}
\centering
\includegraphics[width=0.9\textwidth]{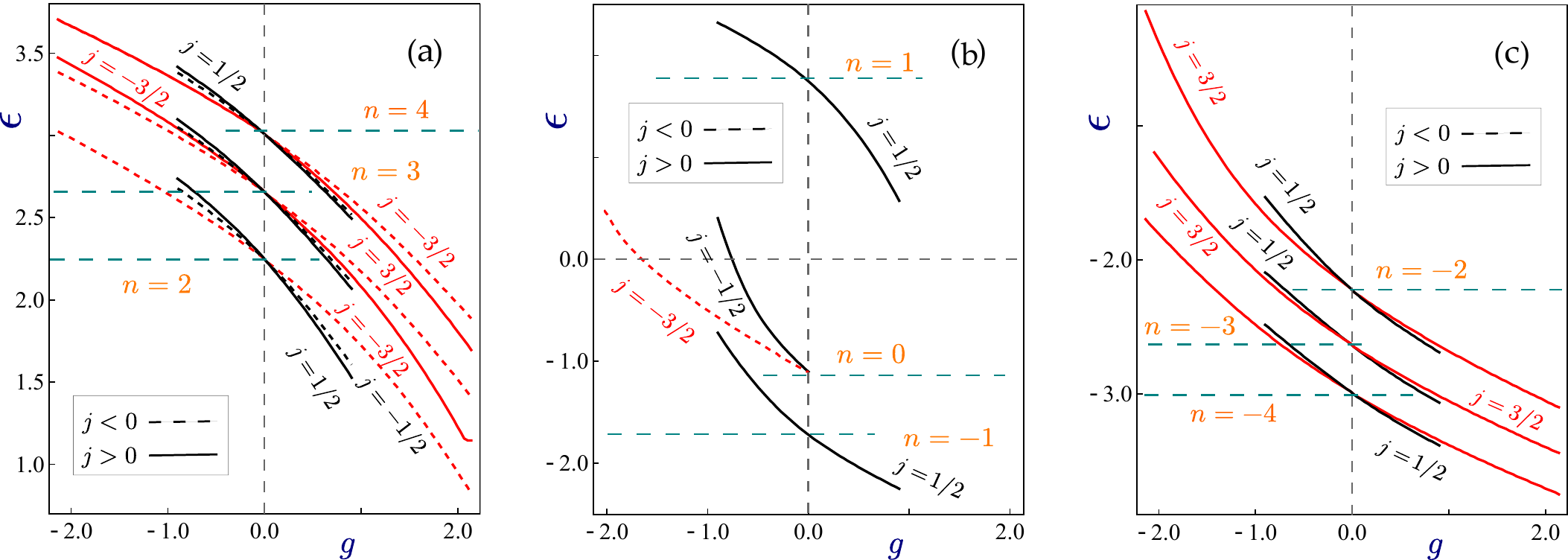}
\caption{(Color online)
The energy levels -- distorted Landau levels -- of graphene with a finite bandgap $\Delta=53$ meV in a uniform magnetic field in the presence of a Coulomb
impurity as a function of its charge $g$ with a finite out-of-plane distance $d_\perp = 2.0$nm to the impurity. Panels (a), (b) and (c) show the energy levels with
quantum numbers $4\ge n\ge-4$ for $j = \pm1/2 $ (black curves) and  $j = \pm3/2$ (red curves) as labeled.  The dotted horizontal lines correspond to the energy
levels $\epsilon(n)={\rm sign}(n)\sqrt{2|n|+1}$ in a magnetic field in the absence of a charged impurity.}
\label{FIG:2}
\end{figure}

\medskip

For our numerical investigation, the main goal is to determine the distortion of the Landau levels induced by a charge impurity, which can be either attractive or repulsive, corresponding
to ($g>0$) and ($g<0$), respectively. We study the case of both unscreened  and screened  Coulomb potentials. We obtain the numerical solutions of equation \eqref{BS:Coulomb+magfield},
finding the best-fit energy (with a minimal difference between the fixed right side and energy-dependent left side) for several fixed levels $n$. This equation cannot be solved analytically and
requires a purely numerical treatment.

\medskip
This type of a numerical problem is rather complicated because we don't know the turning points $\rho_{\pm}$ which are the boundaries of integrals in the equation
\eqref{BS:Coulomb+magfield} and are the roots of $\tilde{p}^2(\rho)=0$. In case of a finite bandgap in graphene, no simple analytical approximation for these
turning points exists. In the case of unscreened Coulomb potential,  analytical solutions of quartic equation determining turning points can be written down,
but the actual expression is so complicated that it literally doesn't make sense to use it in any further calculations. And of course, no analytical solutions for
turning points exist for a screened Coulomb potential. In this situation, non-trivial convoluted functional programming is needed, so that we eventually introduce
functions which depend on the other functions of the considered material and field parameters. An additional term proportional to the band gap in the equation
\eqref{BS:Coulomb+magfield}  could be calculated separately, and it definitely affects the sought Landau levels.
\medskip

\subsection{Unscreened Coulomb charge in a magnetic field}

\subsubsection{Charge dependence of the energy levels for fixed magnetic field.}

As already mentioned above, we consider both positive and negative values of the impurity charge $g$. For $g = 0$, we obviously must obtain the standard
Landau levels. It is convenient to introduce the Landau level index $n$, related to the sign of energy, in terms of each unperturbed energy levels are written as
$\epsilon(n)= {\rm sign}(n)\sqrt{2|n| + 1}$ in chosen units of measurement, when $l=\lambda $. This also helps to resolve a serious issue: finding the range
of energies for which we should be looking for our WKB solutions.

\par
Our numerical results for the energy levels -- distorted Landau levels -- of graphene with a finite bandgap are presented in Fig.~\ref{FIG:1}. Here, we consider in-plane impurity
with Coulomb potential $\backsim g/r$. We demonstrate the results for positive, negative, and zero quantum number $n$ on three different plots $(a)$, $(b)$ and $(c)$ of Fig.~\ref{FIG:1}.
As can be seen,  the Landau levels decrease monotonically with increasing $g$ for both positive and negative values of $g$.  The results for a given level are different for various
values of angular momentum $j$ -- the Coulomb interaction lifts the degeneracy of Landau levels, the fact very known in the literature. Also, we have to cut off
the values of $|g|$ around $|j|$, meaning that the range of the charge $g$  is different for the different curves presented in Fig.~\ref{FIG:1}. The condition $\vert g \vert \leq |j|$ is the
 necessary condition to avoid the atomic collapse phenomenon for corresponding levels in the case of pure Coulomb interaction, the larger charges require a separate investigation with
 an adaptation of the WKB approach.

We see that for smaller values of $|j|$, such as $j=1/2$, the change in the energy level due to the electric charge is larger than for larger values of $j$, for the same charge $g$ .
Importantly, we notice from panel $(b)$ that the solutions for $n=0$ exist  inside the gap for the energies $-\Delta < \epsilon < \Delta$  and chosen magnetic field $B=4.3$ T
only for negative charges $g<0$.  This is  because for $g>0$, we obtain only negative solutions for $r_{\pm}$, which is impossible since this is a radial component of the polar
coordinates, and solutions with $r<0$ must be omitted. Coulomb levels, which exist within the gap in the absence of a magnetic field, increase in energy as the field strength
increases and emerge from the gap (see the subsection \ref{magnfield_dependence} and  Fig.\ref{FIG:3}(e)).

\medskip
We also notice an interesting non-linear dependence of the considered Landau levels on $g$ with certain very specific symmetries. For the positive Landau levels $n>0$,
the magnitude of the impurity-induced shift of the energy levels becomes larger with increasing $g$ for $g>0$ and is reduced for $g<0$. For the negative Landau levels
with $n<0$ and $\epsilon < 0$,  the situation is the opposite: it increases for negative $g$ and decreases for $g>0$. This unusual, symmetric dependence could not be
obtained from any analytical approximation; it can only be determined by means of numerical calculations.

\medskip
Fig.~\ref{FIG:2} demonstrates similar results for graphene with a finite gap, but for a finite out-of-plane distance $d_\perp$ so that the Coulomb potential becomes
$V(r) = g/\sqrt{r^2+d_\perp^2}$. For the used value $d_\perp =2$ nm it has a limited effect on the Coulomb interaction, resulting in a slower rate of fading off of
$V(r)$ with increasing $r$. The Landau levels are found to be less distorted and closer to the horizontal levels related to $g=0$, but the key difference is that we do not
need to impose the cutoff  $\vert g \vert =| j| $ since the singularity $r=0$ is screened in this case. We can extend our calculations to obtain results for larger values of $g$ and
investigate the Landau levels for a stronger Coulomb potential, as shown in Fig.~\ref{FIG:2}.

 Results for the Landau levels in the presence of  magnetic field and a subcritical Coulomb  impurity
 were also obtained in Ref.\cite{Sun2014} by using the coupled series expansion method to solve numerically the Dirac equation. Their results in this particular case are in agreement with ours.
 Specifically, the deformation and curvature of the Landau levels are different for positive and negative $g$, which  we also observe in our figures. Apart from this, the magnetic field dependence
 of the obtained Landau levels is also in good agreement with our results presented in Fig.~\ref{FIG:3}. We therefore conclude that the WKB approach yields good results for the problem under
 consideration, even for the lowest energy levels.

\subsubsection{Magnetic field-dependence of the energy levels for fixed Coulomb charges.}
\label{magnfield_dependence}

\begin{figure}
\centering
\includegraphics[width=0.55\textwidth]{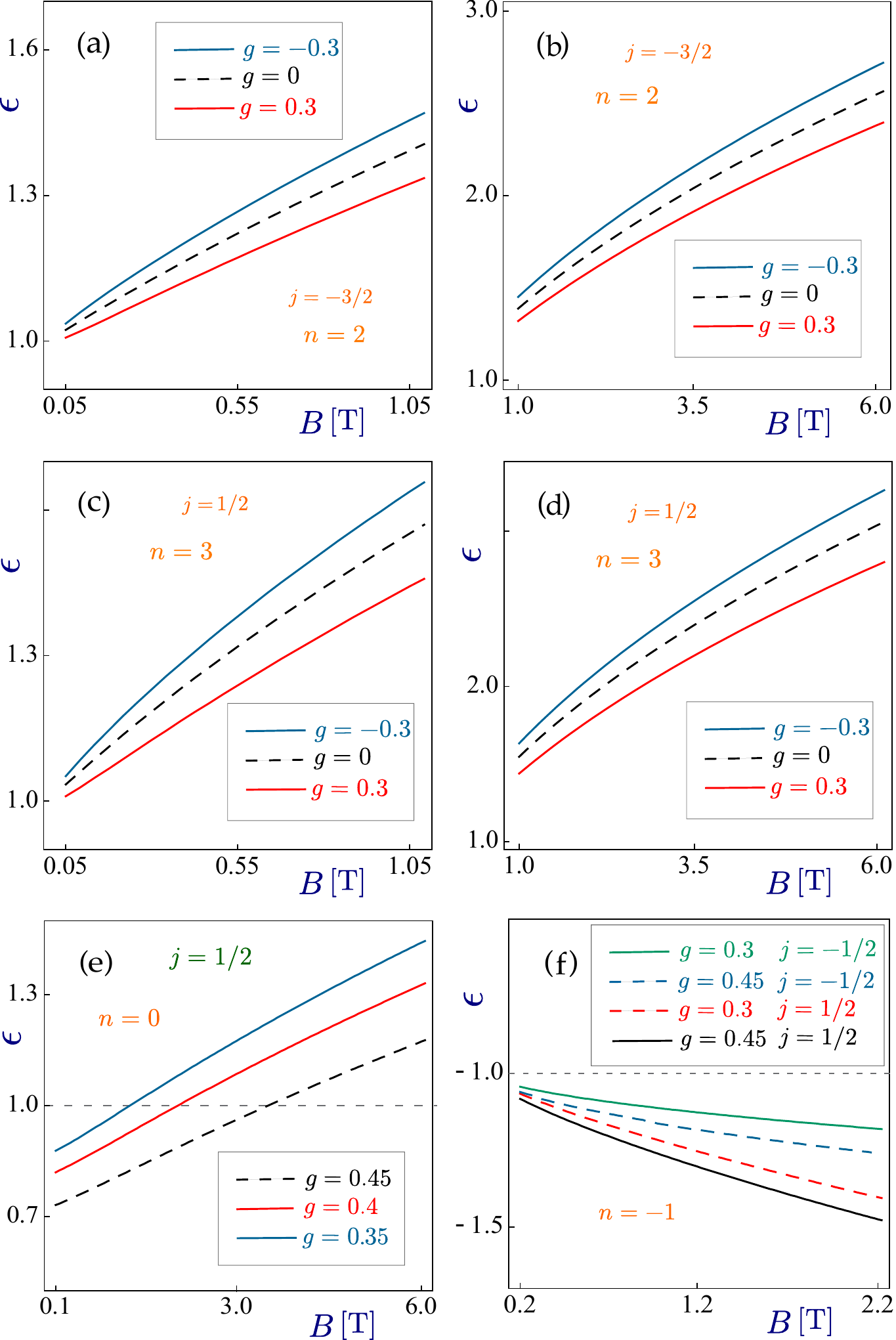}
\caption{(Color online)
The energy levels -- distorted Landau levels -- of graphene with the gap parameter $\Delta=53$ meV in a uniform magnetic field in the presence
of a Coulomb impurity as a function of magnetic field $B$ in $T$ with $d_\perp=0$. In each panel, a blue curve corresponds to an impurity charge $g=-0.3$ (repulsive
potential), black line - to $g=0$, and red one - $g= 0.3$  (attractive potential).  The two upper panels $(a)$ and $(b)$ are related to a state with $n=2$ and $j=-3/2$, while
the lower ones $(c)$ and $(d)$ - to $n=3$ and $j=1/2$. The two left panels $(a)$ and $(c)$ describe a small field $B<1$T, while the right plots $(b)$ and $(d)$ demonstrate
the energy levels for $1.0$T$<B<6.0$T. Panel $(e)$ describes the $n=0$ Landau levels as a function of magnetic field for different values of impurity charge $g$. As we
clearly see, these Landau levels cross $\epsilon = 1$ level and are partially inside the band gap. Panel $(f)$ demonstrates the Landau levels for $n=-1$ for different charges
$g$ and $j=\pm1/2$, as labeled.
}
\label{FIG:3}
\end{figure}

\medskip
Next, we investigate the magnetic-field dependence of the Landau level spectrum in the presence of a Coulomb charge impurity and present our results in Fig.~\ref{FIG:3}.  As before,
we use the length unit $l_0=\lambda_0$, corresponding to the fixed gap parameter $\Delta=53$ meV or the magnetic field $B=4.3$T, to make quantities dimensionless.  In this case
$\epsilon=El_0/\hbar v_F =E/\Delta$ and $\epsilon(n)={\rm sign}(n)\sqrt{2|n|(\lambda/l)^2+1}$  for the levels in a magnetic field without a charged impurity.

We see that for all cases, the Landau levels retain the dependence  $\sim\sqrt{2|n|\hbar v^2_F|eB|/(c\Delta^2) +1}$ expected for gapped graphene in a magnetic field.  The Coulomb impurity
leads to a pronounced shift of the energy levels, with the magnitude of the shift remaining nearly constant throughout the considered magnetic-field range. Consequently, the impurity modifies
the absolute energies of the Landau levels without significantly affecting their magnetic-field  scaling behavior.
\par
The panels (a)-(d) in Fig.~\ref{FIG:3} show that in the presence of an attractive Coulomb impurity $(g > 0)$, the Landau-level energies are always lower than those for the absence of
an impurity $(g = 0)$. In contrast, a repulsive Coulomb potential shifts $(g < 0)$ the energies upward, producing the highest energy levels among the cases considered. The energy
separation between the attractive and repulsive impurity cases increases with increasing magnetic field strength.
\par
We consider magnetic fields with a magnitude of up to $B = 6$ T, which includes the value we used in our previous calculations with a fixed magnetic field, and the limit of very
weak magnetic fields. The low-field regime is examined separately (see the right panels $(a)$ and $(c)$ of Fig.~\ref{FIG:3}) in order to verify the consistency of our numerical approach with
the zero-field limit. Since the Coulomb impurity problem in the absence of a magnetic field has been solved analytically for the attractive potential, it provides a useful benchmark for our
calculations. Our numerical results exhibit excellent agreement with the analytical solution given by Eq.~\eqref{Coulomb-energies}, demonstrating the accuracy and reliability of the method
in the weak-field limit.

\medskip

\begin{figure}
\centering
\includegraphics[width=0.99\textwidth]{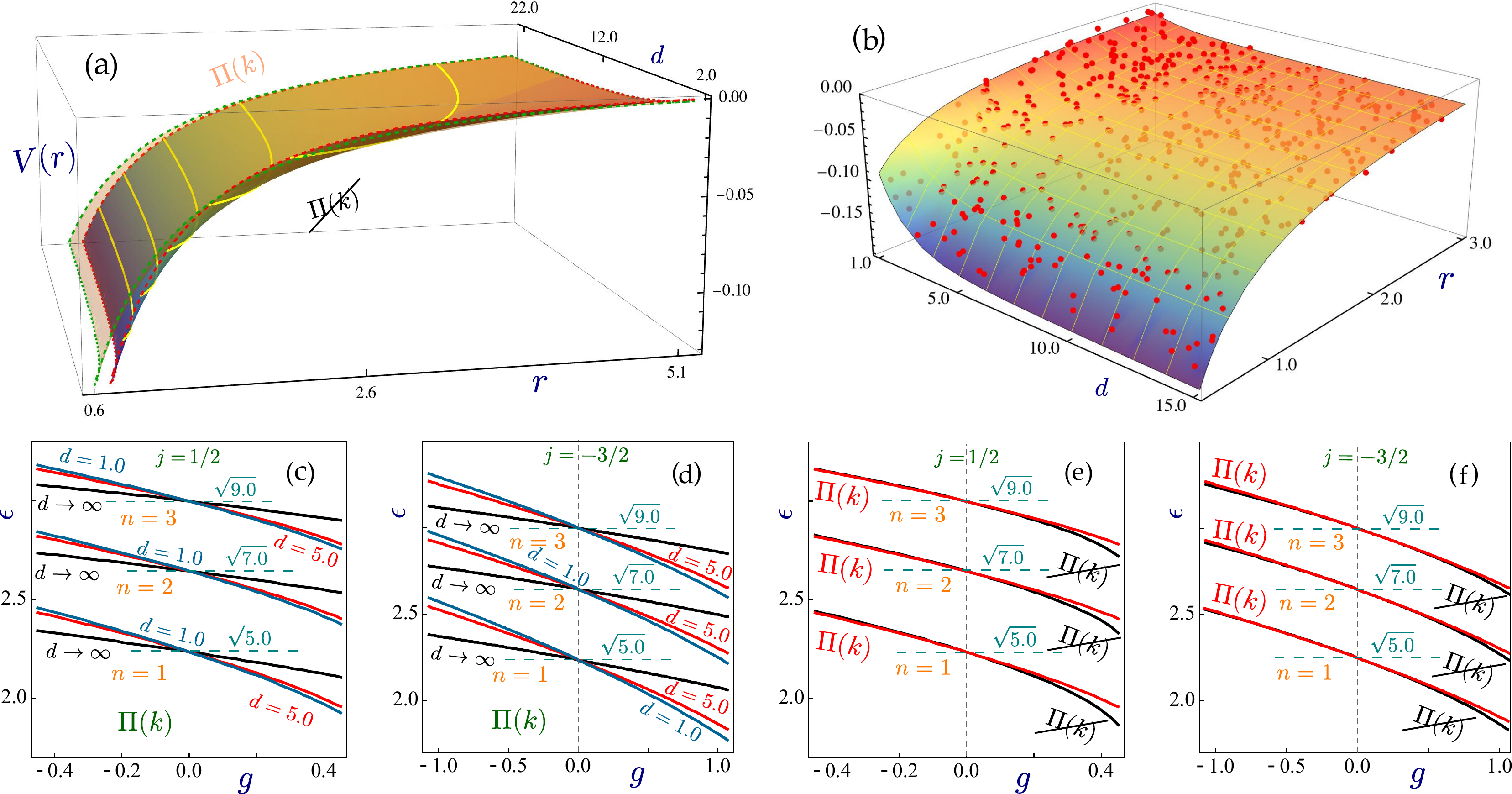}
\caption{(Color online) $(a)$ Influence of polarization on the screened potential obtained from Eq.~\eqref{pot-two_gates_equal}. The potential in the presence of polarization is 
shown by the transparent yellow surface. The magnitude of the potential is reduced when polarization is included. In both cases, the potential is presented as a function of the 
radial distance $r$ and the gate distance $d$ (in units of $l_0$). $(b)$ Verifications and validation of the fitting model given by Eq.\eqref{fitP01}. The rainbow-colored 
three-dimensional surface represents the potential obtained from the analytical fitting formula in equation \eqref{fitP01}.  Panel $(c)$ shows the distorted Landau levels of 
graphene with the gap parameter $\Delta=53$ meV in a uniform magnetic field $B=4.3$T  as a function of a impurity charge $g$ for a screened potential given by Eq.~\eqref{fitP01}. 
Both positive and negative charges of the impurity are considered ($g<0$ on the left side of each plot, $g>0$ - on the right side). Panel $(c)$ corresponds to $j = 1/2 $ and plot 
$(d)$ - $j = - 3/2$, and the results are shown for  distance $d \rightarrow \infty$ (black curves, equivalent to the unscreened Coulomb potential), $d = 5.0$ (red curves) and $d = 1.0$ 
(blue curves), as labeled. Panels $(e)$ and $(f)$ demonstrate the effect of polarization on the Landau levels in the screened potential  for the same levels as  in panels $(c)$ and $(d)$
 but addressed for the distance $d=5.0$ (symbol $\cancel{\Pi(k)}$  means that the polarization is not taken into account).
  }
\label{FIG:4}
\end{figure}
\medskip

\begin{figure}
\centering
\includegraphics[width=0.49\textwidth]{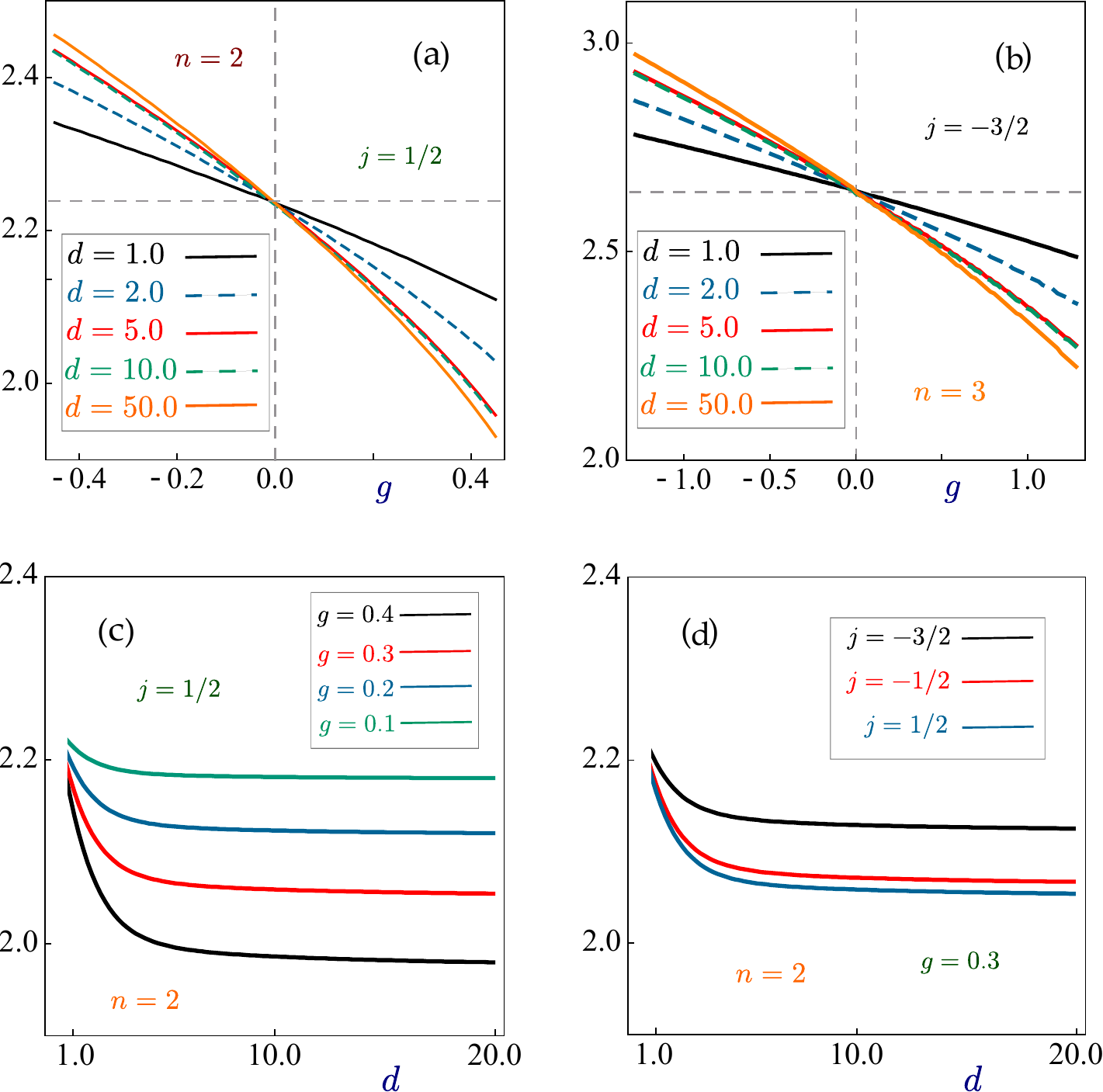}
\caption{(Color online) Panel $(a)$ shows the distorted Landau levels of graphene with the gap parameter $\Delta=53$ meV in a uniform magnetic field $B=4.3$T  
as a function of a impurity charge $g$ for $n=2$ and $j=1/2$ in the screened potential \eqref{fitP01}. Both positive and negative charges of the impurity are considered 
($g<0$ on the left side of each plot, $g>0$ - on the right side). Panel $(b)$ shows a similar dependence for  $n=3$ and $j=-3/2$. In both panels $(a)$ and $(b)$, each curve 
corresponds to a set of multiple distances $d=1.0$ (black curve), $3.0$ (blue dashed curve), $5.0$ (red curve), $10.0$ (green dashed curve) and $50.0$ (orange curve), as labeled. 
Panel $(c)$ demonstrates the dependence of the distorted Landau levels on the distance $d$ for $n=2$ and $j=1/2$. Each curve corresponds to a different impurity charge 
$g=0.4$, $0.3$, $0.2$ and $0.1$, represented by the black, red, blue, and green curves, respectively. Panel $(d)$ shows a similar dependence of the distorted Landau levels on $d$ for 
$n=2$ and $g=0.3$, where each curve corresponds to a different value of $j$: $j=1/2$ (blue), $j=-1/2$ (red) and $j=-3/2$  (black). In all calculations and all panels, the polarization 
$\Pi(k)$ has been included into consideration.}
\label{FIG:5}
\end{figure}
\medskip
\subsection{Screened Coulomb potential in a magnetic field}

For the screened potential, we used the  analytical fitting expression given by Eq.~\eqref{fitP01}. Importantly, both the impurity potential and its spatial derivative are required
to determine the Landau levels in the presence of a charged impurity.  In the absence of polarization, the potential becomes equivalent to the Coulomb potential, $\backsim g/r$,
for large gate distance $d$. This is exactly what we require from our analytical fitting model. In principle, we are interested in determining how strongly the screening of the
Coulomb potential, due to polarization and closely situated gates, affects the actual results. We find that, in general, the effect of polarization is not very large, and it becomes
more pronounced at very small $r$ and for larger $d$, where the potential approaches the Coulomb limit, as specifically demonstrated in panel $(a)$ of our Fig.~\ref{FIG:4}.
The absence of polarization $\Pi(k)$ is marked in  Fig.~\ref{FIG:4} as  $\cancel{\Pi(k)}$. The numerical computations shown in Fig.~\ref{FIG:4} $(b)$ indicate that the analytical
fitting \eqref{fitP01} is in  a good agreement with the general expression given by Eq.~\eqref{pot-two_gates_equal}.

 Our numerical results for the Landau levels in the presence of a screened Coulomb potential are presented in panels $(c)$-$(f)$ of Fig.~\ref{FIG:4}.  First, we observe that the
 effect of the screened potential is weaker than that of the unscreened Coulomb potential for the same value of $g$. This behavior is expected, since screening reduces the effective
  strength of the impurity potential. As a result, the deviation of the Landau-level energies from their impurity-free values (regular Landau levels) is smaller for finite screening.
  The limiting case $d \rightarrow \infty$ corresponds to the regular unscreened Coulomb potential. Although the overall dependence of the energy on the impurity strength $g$
  remains qualitatively similar, with the energy decreasing as $g$ increases for both positive and negative values, the symmetry of this dependence is very different in contrast to
  the regular Coulomb potential, where the behavior is much more symmetric with respect to the sign of $g$.

\par
The screened potential \eqref{fitP01} depends on the impurity strength $g$ and the  distance $d$ of graphene from gates, and, given the form of the potential itself, numerical
calculations become considerably more complicated than in the case of a purely Coulomb potential.  Therefore, we restricted our consideration in Fig.~\ref{FIG:4} to momenta
$j = \pm 1/2$ and $\pm 3/2$, and to several values of screening distances $d$ (less $d$ means a stronger screening effect).

\par
In panels (c)--(f) of Fig.~\ref{FIG:4}, we examine how the polarization function affects the actual Landau levels. We consider two different fitting models: one that accounts for polarization
 and one that does not. We then calculate the Landau levels separately for both cases. We find that the effect of polarization is generally small, and the qualitative behavior remains the same.
However, for large $g$, the Landau levels are distorted significantly more when the polarization function is neglected, and the general distortion (for both absent and present polarization)
 becomes noticeable. In contrast, the solutions obtained close to the regular Landau levels remain nearly unchanged.

\par
Finally, since the screening potential, its modeling, and the resulting Landau levels are central to this work and are of particular interest for experimental studies, we would like to discuss the
dependence on the screening distance $d$ in more detail. First, we plot the dependence of the distorted energy levels on the impurity charge $g$ for a wide range of screening distances $d$ and
present the results in panels (a) and (b) of Fig.~\ref{FIG:5}. We observe that, for larger $d$, the distortion and curvature of the Landau levels become more pronounced for the same impurity charge
$g$, since the potential approaches the standard Coulomb form, while the polarization $\Pi(k)$ is still taken into account. We also note that the changes in the Landau levels are most significant
 at small $d$, whereas they become progressively less noticeable as $d$ increases. We further show the explicit dependence on $d$ in panels $(c)$ and $(d)$ of Fig.~\ref{FIG:5}. As $d$ increases,
 the energy levels approach constant values and their dependence on $d$ becomes weak. A significant dependence on $d$ is observed at distances $1<d<4 $ (actual distances $10$ -- $50$ nm),
  and the effect of the  presence of gates at such small distances must undoubtedly be taken into account (we do not consider distances between the gates and the sample of less than $10$ nm,
which are apparently impracticable).

 The modification of the Landau levels is also more significant for larger impurity charges $g$, as expected.
 In addition, we find that the levels with $j=1/2$ are most strongly affected by the screening distance $d$, whereas those with $j=-3/2$ are least affected.

\section{Concluding remarks}
\label{sec5}

In this paper, we presented a detailed theoretical and numerical analysis of the electron energy levels in gapped graphene subjected to a magnetic field  in the presence of
a charged impurity, using the Wentzel–Kramers–Brillouin approximation. This method is particularly useful for quantum-mechanical problems for which exact analytical solutions
are not available, including the determination of bound-state energy levels. The latter problem is the focus of the present work.

\medskip
We derived analytical expressions for the WKB wave functions of electrons in gapped graphene in the presence of a magnetic field and a charged impurity. We also obtained
a Bohr–Sommerfeld-type quantization condition using the complex Wentzel method, which is based on the argument principle, and demonstrated the role of an additional
geometric-phase term in this condition. By solving the quantization condition numerically, we examined two important limiting cases: vanishing Coulomb interaction, which
recovers the conventional Landau levels, and vanishing magnetic field, which yields the Coulomb energy levels. Our numerical results agree with these two well-known limiting
cases within the numerical precision.

\medskip
Our numerical results reveal some interesting features of the distorted Landau levels, including their dependence on the impurity charge $g$, for both positive and negative charges,
and on the magnetic field strength. In particular, we identified symmetry properties of the Landau-level spectra:  the curvature has opposite signs for positive and negative Landau
levels, as seen in all the levels we calculated.

 Special attention was paid to the  screening of the Coulomb potential due to polarization effects and  the presence of metallic gates at some distances from a sample, the case which
  is usually present in experiments. It is worth emphasising that the calculation of the WKB integrals for the screened potential, defined by an integral expression, required an analytical
  approximation to the potential in order to determine the classical turning points accurately. While polarization effects have little influence on the energy spectrum for sufficiently large
  gap parameters, metallic gates located approximately $10$--$50$ nm from the sample produce significant screening. Our results therefore demonstrate that, when studying the physical
  properties of graphene and other graphene-like structures, it is necessary to  take into account the  influence of nearby metallic gates.

 In this study, we considered a heterostructure consisting of two gates placed at equal distances on either side of the graphene sample, with the same dielectric permittivity in the regions
 between the graphene and the gates. In this configuration, the screened potential is given by Eq. (\ref{pot-two_gates_equal}). One can also consider other configurations, including
 gates situated at different distances from the sample, and heterostructures with different dielectric permittivities in the intermediate  regions,  using the more general expression in
 Eq. (\ref{pot-two_gates_diff}). The latter expression also encompasses the limiting case of a single gate, described by Eq. (\ref{pot-one_gate}).

\medskip
Our results for the energy levels of graphene with screened and unscreened charged impurities in a magnetic field may be relevant to studies of layered Dirac materials, including systems with flat bands,
as well as to investigations of quantum dots.

\begin{acknowledgments}
The authors thank E.V. Gorbar and S.G. Sharapov for discussions and useful suggestions.
A.I.  acknowledges the  support by the funding received from TradB-56-75, PSC-CUNY Award \# 68386-00 56.
The work of V.P.G. was supported by the Swiss NSF within the Ukrainian-Swiss Joint research project “Transport and thermodynamic phenomena
in low-dimensional materials with flat bands” (Grant No. IZURZ2$\_224624$).
\end{acknowledgments}
\medskip

\appendix

\section{Limiting case: Zero magnetic field but nonzero Coulomb potential. }
\label{apaa}

\subsection{Zero magnetic field but nonzero Coulomb potential}

Let us show  that in the case of attractive Coulomb potential $V(r)=-Ze^2/(\kappa r)=-\hbar v_F g/r$ without magnetic field the WKB quantization condition (\ref{WKB-rule-Delta})
gives energy levels coinciding with known exact solutions of ($2+1$)-dimensional Dirac equation. The integral with arctan vanishes in this case and for the second integral in the left hand side of
Eq. ~(\ref{WKB-rule-Delta}) we have
\begin{align}
I=\frac{\Delta}{2}\int\limits_{r_-}^{r_+}\frac{dr V'(r)a(r)}{p(r)[(E-V(r))^2-\Delta^2]}=\frac{\Delta g(\hbar v_F)^2 j}{2}\int\limits_{r_-}^{r_+}
\frac{dr}{\sqrt{(E r+g\hbar v_F)^2-\Delta^2r^2-(\hbar v_F j)^2}[(E r+g\hbar v_F)^2-\Delta^2r^2]}.
\end{align}
The roots of the equation $p^2(r)=0$ are
\begin{align}
r_{\pm}=\frac{\hbar v_F}{\Delta(1-\epsilon^2)}\left[\alpha\epsilon\pm\sqrt{g^2\epsilon^2-(j^2-g^2)(1-\epsilon^2)}\right],
\quad \epsilon=E/\Delta,\quad |\epsilon|<1.
\end{align}
Writing
\begin{align}
(E r+g\hbar v_F))^2-\Delta^2r^2-(\hbar v_F j)^2=(\Delta^2-E^2)(r-r_{-})(r_+ -r),
\end{align}
we arrive at the expression
\begin{align}
I=\frac{\Delta g(\hbar v_F)^2 j}{2(\Delta^2-E^2)^{3/2}}\int\limits_{r_-}^{r_+}\frac{dr}{\sqrt{(r-r_{-})(r_+ -r)}[(r-r_{-})(r_+ -r)+R^2]},\quad
R^2=\frac{(\hbar v_F j)^2}{\Delta^2-E^2}.
\end{align}
Introducing dimensionless variable $r=(\hbar v_F/\Delta)x$ we rewrite the last expression in the form
\begin{align}
I&=\frac{g j}{2(1-\epsilon^2)^{3/2}}\int\limits_{x_1}^{x_2}\frac{dx}{\sqrt{(x-x_1)(x_2-x)}[(x-x_1)(x_2-x)+\frac{j^2}{1-\epsilon^2}]}\nonumber\\
&=\frac{g j}{2(1-\epsilon^2)^{3/2}(x_2-x_1)^2}\int\limits_{0}^{1}\frac{dy}{\sqrt{y(1-y)}[y(1-y)+b^2]},\qquad b^2=\frac{j^2}{(1-\epsilon^2)(x_2-x_1)^2},
\end{align}
where we made the change $x=(x_2-x_1)y+x_1$ and $x_1, x_2$ are
\begin{align}
x_{1,2}=\frac{1}{1-\epsilon^2}\left[g\epsilon\mp\sqrt{g^2\epsilon^2-(j^2-g^2)(1-\epsilon^2)}\right].
\end{align}
The integral is evaluated exactly (for example,  using Eq. 3.121.2 in \cite{Gradshteyn})
\begin{align}
&\int\limits_{0}^{1}\frac{dy}{\sqrt{y(1-y)}[y(1-y)+b^2]}=\frac{2\pi}{|b|\sqrt{1+4b^2}}.
\end{align}
Finally, we get
\begin{align}
I=\frac{g j}{2(1-\epsilon^2)^{3/2}(x_2-x_1)^2}\frac{2\pi}{|b|\sqrt{1+4b^2}}=(\pi/2){\rm sign}(j).
\end{align}
The WKB quantization condition (\ref{WKB-rule-Delta}) takes the form
\begin{align}
\sqrt{1-\epsilon^2}\int\limits_{x_1}^{x_2}\frac{dx}{x}\sqrt{(x-x_1)(x_2-x)}=\pi\left(n+\frac{1-{\rm sign}(j)}{2}\right),\qquad n=0,1,\dots\,.
\label{quantization-Coulomb}
\end{align}
The integral is equal to
\begin{align}
\frac{\pi(x_1+x_2-2\sqrt{x_1x_2})}{2}=\pi\left(\frac{g\epsilon}{1-\epsilon^2}-\frac{\sqrt{j^2-g^2}}{\sqrt{1-\epsilon^2}}\right).
\label{WKB-Coulomb-integral}
\end{align}
Finally, we find from Eqs.(\ref{quantization-Coulomb}) and (\ref{WKB-Coulomb-integral})  the energy levels in the Coulomb potential given by Eq.(\ref{Coulomb-energies}) in the main text.

\section{Deriving some useful relations for the screened Coulomb potential}
\label{apb}

In general the potential can be written in the form
\begin{align}
V(r)=-\frac{e q}{\kappa r}f\left(\frac{d}{r},\frac{\lambda}{r}\right),
\end{align}
where function $f$ is a dimensionless function of two dimensionless variables.

Let us first consider the case when the polarization is absent which corresponds to the case of large gaps ($\Delta\to\infty$), then the interaction of electrons with impurity of
a charge $q$ potential is given by the expression
\begin{align}
V(r)=-\frac{e q}{\kappa}\int\limits_0^\infty dk \tanh(kd) J_0(k r).
\label{pot-ero_poparization}
\end{align}
To evaluate the  integral we consider two series for $\tanh x$:
\begin{align}
&\tanh x=1+2\sum\limits_{n=1}^\infty (-1)^n e^{-2n x},\\
&\tanh x=8x\sum\limits_{n=1}^\infty \frac{1}{\pi^2(2n-1)^2+4x^2}.
\end{align}
We get, respectively
\begin{align}\label{series-1}
&V(r)=-\frac{e q}{\kappa}\left[\frac{1}{r}+\frac{1}{d}\sum\limits_{n=1}^\infty \frac{(-1)^n}{\sqrt{n^2+(r/2d)^2}}\right],\\
&V(r)=-\frac{2e q}{\kappa d}\sum\limits_{n=1}^\infty K_0\left(\pi(n-1/2)\frac{r}{d}\right).
\label{series-exponentials}
\end{align}
The first expression is suitable for finding asymptotics for $r\ll d$, while the second for $r\gg d$,
\begin{align}
&V(r)\simeq-\frac{e q}{\kappa}\left[\frac{1}{r}-\frac{\ln 2}{d}\right],\qquad r\ll d,\\
&V(r)\simeq-\frac{e q}{\kappa}\frac{2}{\sqrt{r d}}e^{-\pi r/2d}, \qquad r\gg d.
\end{align}

A  closed form can be obtained from Eq.(\ref{series-exponentials}) if we use the representation for the function $K_0(z)$:
\begin{align}
K_0(z)=\int_0^\infty e^{-z\cosh t}dt,\qquad Re\,z>0.
\end{align}
Summing over $n$ we get
\begin{align}
V(r)=-\frac{e q}{\kappa d}\int_0^\infty\frac{dt}{\sinh\left(\frac{\pi r}{2d}\cosh t\right)}.
\end{align}
This expression is convenient for obtaining numerically the dependences of the potential $V(r)$ on $r$ and $d$.


\end{document}